\documentstyle[prb,aps]{revtex}
\begin{document}
\wideabs{
\title{Spin gap in low-dimensional Mott insulators with orbital degeneracy}
\author{L. Guidoni$^{1,2}$, G. Santoro$^{1,2}$, S. Sorella$^{1,2}$,
A. Parola$^{1,3}$, and E. Tosatti$^{1,2,4}$}
\address{$^{(1)}$ Istituto Nazionale per la Fisica della Materia (INFM),
Italy\\
$^{(2)}$ International School for Advanced Studies (SISSA), Via Beirut 4,
Trieste, Italy\\
$^{(3)}$ Istituto di Scienze Fisiche,
Universit\`a di Milano, via Lucini 3, Como, Italy\\
$^{(4)}$ International Center for Theoretical Physics (ICTP), Strada Costiera,
Trieste, Italy}
\date{\today}
\maketitle
\begin{abstract}
We consider the exchange Hamiltonian 
$H_{\rm ST} = - J \sum_{<rr'>} 
( 2 {\bf S}_r \cdot {\bf S}_{r'} - \frac{1}{2}) 
( 2 {\bf T}_r \cdot {\bf T}_{r'} - \frac{1}{2}) \;,$
describing two isotropic spin-1/2 Heisenberg antiferromagnets coupled by a 
quartic term on equivalent bonds.
The model is relevant for systems with orbital degeneracy and strong 
electron-vibron coupling in the large Hubbard repulsion limit. 
To investigate the ground state properties we use a Green's Function 
Monte Carlo, calculating energy gaps and correlation functions, the latter
through the forward walking technique.  
In one dimension we find that the ground state is a ``crystal'' of valence 
bond dimers. 
In two dimensions, the spin gap appears to remain finite in the thermodynamic 
limit, and, consistently, the staggered magnetization -- signal of N\'eel long
range order -- seems to vanish. 
From the analysis of dimer-dimer correlation functions, however, we find 
no sign of a valence bond crystal. 
A spin liquid appears as a plausible scenario compatible with our findings. 
\end{abstract}
%
%
}		 
\newpage

The object of this study is a particular spin-exchange Hamiltonian in which
two spin-1/2 variables -- a standard spin ${\bf S}$, and a pseudo-spin
${\bf T}$ representing an orbital degree of freedom -- are coupled
together in the following way:
\begin{equation} \label{hst:eqn}
H_{\rm ST} = - J \sum_{<rr'>} 
( 2 {\bf S}_r \cdot {\bf S}_{r'} - \frac{1}{2}) 
( 2 {\bf T}_r \cdot {\bf T}_{r'} - \frac{1}{2}) \;,
\end{equation}
the lattice summation running over the nearest neighbor sites of a lattice.
The symmetry group displayed by Eq.\ (\ref{hst:eqn}) includes 
$SU(4)$.\cite{Santoro}
This Hamiltonian was shown to describe the low-energy physics of an insulating
crystal with one-electron per site in a two-fold degenerate orbital, 
in the limit of large on-site repulsion (Mott insulator) and in presence of
Jahn-Teller effect.\cite{Santoro}
The derivation of Eq.\ (\ref{hst:eqn}), henceforth referred to as ST-model, is 
standard.\cite{Santoro,Auerbach,Fabrizio} 
The crucial physical condition to be verified is that, among the possible 
two-particle states obtained upon virtual hopping, the {\em inter-orbital 
singlet\/} should turn out to be the lowest in energy, 
which is indeed the case when a strong dynamical JT effect is at play. 
A different (perhaps more common) physical situation is that the lowest
two-particle intermediate state is a triplet (as a result of a Hund's rule),
in which case the exchange Hamiltonian has a different form and spin
ferromagnetism is the natural outcome.\cite{KK,Auerbach}
In the general case, the exchange Hamiltonian contains pseudo-spin anisotropic
terms, and its phase diagram has been addressed with different techniques in
Refs.\ \cite{Auerbach,Last_Khomskii}

$C_{60}$ charge transfer compounds, based on $C_{60}^-$ ions,\cite{C60:rev} 
or molecular compounds with similar characteristics, 
are potential candidates to the realization of a scenario in which
the Hamiltonian in Eq.\ (\ref{hst:eqn}) is possibly relevant. 
$C_{60}$ has a triply degenerate $t_{1u}$ molecular orbital, coupled 
to several intramolecular vibrations, resulting in an important 
dynamical JT effect.\cite{AMT} 
The electron-phonon coupling leads to a substantial pairing energy 
$\approx 0.1 eV$ which is, however, overwon by a substantially larger 
Hubbard $U\approx 1-1.5 eV$,\cite{Gunnarsson} resulting in a Mott insulating 
behavior. 
The lattices tend to have anisotropic lattice constants, with a
pronounced quasi-one-dimensional character.\cite{C60:rev} 
As for the virtual intermediate states of $C_{60}^{2-}$, singlets and triplet
are close in energy, at least in solutions.\cite{C60:exp} 
If the triplet prevails, as is possibly the case in
TDAE-$C_{60}$,\cite{Auerbach,TDAE:nota} there can be spin ferromagnetism. 
We believe, however, there is room in some future system for the alternative 
possibility of a spin gap, and the formation of the spin-orbital VB state 
described here.

%
Simple variational arguments are quite instructive in elucidating, as a 
function of the dimensionality of the lattice, the physical role played by 
the quadratic and quartic terms in which the ST-model can be rewritten, 
$H_{\rm ST} = J \sum_{<rr'>} [{\bf S}_r \cdot {\bf S}_{r'} + 
{\bf T}_r \cdot {\bf T}_{r'}  - 
4 ({\bf S}_r \cdot {\bf S}_{r'}) \, ({\bf T}_r \cdot {\bf T}_{r'})]$. 
The quadratic terms are standard antiferromagnetic couplings, which, alone,
would drive the system to a N\'eel ordered state for both spin and orbital
variables. 
The quartic term, on the contrary, gains from large values of 
$\langle {\bf S}_r \cdot{\bf S}_{r'}\rangle$ and  
$\langle {\bf T}_r \cdot{\bf T}_{r'}\rangle$: the optimal situation is a
singlet valence bond (VB) state for both S and T variables, where
$\langle {\bf S}_r \cdot {\bf S}_{r'} \rangle = 
\langle {\bf T}_r \cdot {\bf T}_{r'} \rangle = -3/4$. 
Let us compare the energy of the following two states: 
(i) The product of two independent Heisenberg ground states,
$|\Psi_{\rm 2-H}\rangle = |\Psi^{(S)}_{\rm H}\rangle \otimes
                          |\Psi^{(T)}_{\rm H}\rangle$, 
and (ii) the product of two VB states,  
$|\Psi_{\rm 2-VB}\rangle = |\Psi^{(S)}_{\rm VB}\rangle \otimes
                           |\Psi^{(T)}_{\rm VB}\rangle$ 
where each VB state has the simple form of a product of dimers on 
adjacent sites. 
In $D=1$ we take 
$|\Psi_{\rm VB}\rangle = (1 2) \, (3 4) \, \cdots \, (L-1 \; L)$,
where $(ij)$ denotes a singlet between sites $i$ and $j$. 
In $D=2$ we consider any state with nearest-neighbor 
pairs coupled into singlets. 
The energy contribution of the quartic term is substantially lower for the 
VB state, $-4J (1/2)(-3/4)^2$, than for the Heisenberg state for $D\le 2$,
$-4J D(\epsilon_{\rm H}/DJ)^2$, where  
$\epsilon_{\rm H}/J=D\langle \Psi^{(S)}_{\rm H}|
{\bf S}_r \cdot {\bf S}_{r'} |\Psi^{(S)}_{\rm H} \rangle \approx -0.4431$,
in $D=1$, or $\approx -0.66$, in $D=2$.  
The energy per site of the two states including the quadratic terms are:
$\epsilon_{\rm 2-H}=2\epsilon_{\rm H} -4J D(\epsilon_{\rm H}/DJ)^2$, and  
$\epsilon_{\rm 2-VB}=2\epsilon_{\rm VB} -4J (1/2)(-3/4)^2$ 
where $\epsilon_{\rm VB}=(-3/8)J$ is the energy per site of the VB state. 
In $D=1$ the VB state wins over the Heisenberg state. 
In $D=2$, the previous crude variational estimate would give the 
Heisenberg state as favored. 
The VB state considered here is, however, very poor: 
its energy per site, neglecting the quartic term, is only $-0.375 J$, 
whereas it is well known that resonating VB states close in energy to 
the Heisenberg ground state can be constructed in $D=2$.\cite{Doucot}

To show that the VB scenario is correct in 1D, we calculate gaps and
correlation functions using exact Lanczos diagonalizations of chains up to 
$14$ sites, and Green Function Monte Carlo (GFMC) for longer chains 
employing the bias-correction scheme of Calandra and Sorella.\cite{Calandra}
The trial wavefunction we have used is a Jastrow-type function 
$\langle R | \psi_{\rm trial} \rangle = \langle R | \exp \{ -\sum_{(rr')}
[v_{rr'} (S^z_r S^z_{r'} + T^z_r T^z_{r'}) -4 w_{rr'} ( S^z_r S^z_{r'} ) 
( T^z_r T^z_{r'})]\} | R \rangle $
where $ | R \rangle$ is a configuration in the basis 
in which $S^z_r$ and $T^z_r$ are diagonal.
We allow, in general, long range tails in $v_{rr'}$ and $w_{rr'}$, which are
adjusted through a variational calculation.
Fig.\ \ref{results_1d:fig} shows the finite-size value of the gap between the 
ground state and the first excited state both in the $S^z_{tot}=T^z_{tot}=1$ 
($\Box$) and in the $S^z_{tot}=T^z_{tot}=0$ sectors ($\bigcirc$). 
The full symbols refer to $H_{\rm ST}$, whereas the open symbols refer to
two decoupled Heisenberg chains (no quartic term in Eq.\ \ref{hst:eqn}). 
All these excited states have momentum $\pi$ relative to the ground state.
For the Heisenberg case, the excited state is a triplet and the model has 
gapless excitations with gaps decreasing as $1/L$ for 
$L\rightarrow \infty$ (dashed line).
For $H_{\rm ST}$, the situation is different: The lowest excited state is
a singlet, with a finite size gap $\Delta E(S=0)$ decreasing faster than $1/L$ 
for $L\rightarrow \infty$. 
The triplet, instead, lies above with a gap $\Delta E(S=1)$ extrapolating 
to a finite value as $L\rightarrow \infty$, $\Delta E(S=1) \approx 0.5969$. 
This is a clear signal of a {\em degenerate\/} infinite volume ground 
state with a {\em spontaneous breaking of translational symmetry\/} and a gap
to all excitations, consistent with a VB phase. 
Further support to this interpretation can be gained by direct inspection of
the spin-spin correlations, $\langle S^z_0 S^z_i \rangle$, 
and the dimer-dimer correlations along the chain 
$\langle (S^z_0S^z_{1}) (S^z_rS^z_{r+1}) \rangle$. 
The inset of Fig.\ \ref{results_1d:fig} shows a log-log size 
scaling of the peak of the two structure factors, located at 
momentum $\pi$. 
While the dimer-dimer structure factor peak, 
$\sum_r (-1)^r \langle (S^z_0S^z_{1}) (S^z_rS^z_{r+1}) \rangle$ 
tends to diverge linearly with the length $L$ of the chain, the usual spin-spin 
structure factor tends to a finite value as $L\rightarrow \infty$. 

%
The possibility that a valence bond state might be the ground state of the
ST-model even in 2D was put forward in Ref.\ \cite{Santoro}. 
The argument was based on a $SU(n)$-invariant model, due to Affleck,
\cite{Affleck} which is unitarily equivalent to the ST-model in 
Eq.\ (\ref{hst:eqn}) for $n=4$. 
It turns out that in the $n\to \infty$ limit such a $SU(n)$-invariant
model should be characterized, in $D=2$, by a plaquette resonating VB state, 
\cite{Santoro} as discussed in Ref.\ \cite{Leung}. 
Such a phase has resonating VB plaquettes forming alternating disordered
columns, but breaks translational invariance along the axis perpendicular
to the disordered columns. 
 
%
We are now going to discuss the results we obtain with the GFMC for the 
ST-model on the square lattice. 
Fig.\ \ref{gap2d:fig} shows the finite size gap between the ground state and 
the lowest excited triplet state for the ST-model on a $L\times L$ lattice as 
a function of the inverse volume. 
The data for the ST-model, at variance with those of the Heisenberg model,
suggest the presence of a spin gap in the thermodynamic limit,
although some care should be exercised because of the large error bar on
the largest sample data, and the possibility of a non-trivial size scaling.
The solid line is simply a straight line going through the two data points
for a $6\times 6$ and a $10\times 10$ lattices. The extrapolated value of the
spin gap is $\sim 0.373 \; J$.

In principle, either a VB crystal with broken translational symmetry
or a homogeneous spin liquid is compatible with the existence of a spin
gap. 
In order to investigate more closely the possible kinds of long range order which might
characterize the ground state of the ST-model in $D=2$, 
we have calculated the expectation value of several spin-spin
correlation functions by the forward walking technique.\cite{Calandra}
Fig.\ \ref{sq2d:fig} shows the results for the structure factors
related to both standard spin-spin correlations 
$\sum_{r} (-1)^r \langle S_0^z S_r^z \rangle$ (solid squares) and 
dimer-dimer correlations in $\hat{x}$ direction
$\sum_r e^{iqr} \langle (S^z_0S^z_{0+\hat{x}}) (S^z_rS^z_{r+\hat{x}}) \rangle$ 
for different momenta: $(0,\pi)$ (circles),$(\pi,0)$ (hexagons)
$(\pi,\pi)$ (triangles).
For comparison, we include the results for the $(\pi,\pi)$ spin-spin 
structure factor of the $D=2$ Heisenberg model,  
$\sum_{r} (-1)^r \langle S_0^z S_r^z \rangle$ (open squares). 
We observe that all the dimer-dimer correlations considered show no sign
whatsoever of diverging linearly with the volume, i.e., we find that 
$S_{\rm dim-dim}(q)/L^2 \to 0$. 
This finding rules out the possibility of a crystalline VB state, even in the 
form of a plaquette resonating VB.\cite{Leung} 
On the other hand, the standard $(\pi,\pi)$ spin-spin structure factor 
-- signal of a N\'eel ordered state -- , does not seem either to diverge
linearly with the volume, i.e., $S(\pi,\pi)/L^2$ also seems to go to zero. 
The latter finding is compatible with the presence of a gap suggested by
the results in Fig.\ \ref{gap2d:fig}. 
Summarizing, although further work is necessary for a definite answer to 
the nature of the ST-model ground state on the square lattice, it seems 
fair to conclude that our data suggest a disordered VB state as the most 
likely candidate.

In conclusion, we have studied a particular exchange 
Hamiltonian, Eq.\ (\ref{hst:eqn}), 
which describes the low-energy physics of a Mott insulator with orbital
degeneracy in the regime in which the inter-orbital singlet is the 
lowest-energy intermediate state available to virtual hopping. 
This regime tends to be further favored by a relatively strong 
electron-phonon coupling. 
The basic ingredients needed for a realistic system to be a candidate VB 
described in this work are:
i) orbital degeneracy not trivially removed by cooperative JT and/or
crystal field effects;
ii) relatively large molecules with a strong electron-phonon coupling, 
so as to make the inter-orbital singlet favored as compared to the Hund's 
rule triplet;
iii) narrow bands with relatively large on-site Hubbard $U$, so as to stabilize
a Mott-Hubbard insulator at exactly one electron per site;
iv) reduced dimensionality. 
At variance with the standard Hund's rule case (the high spin two-particle state
has lowest energy), which would instead favor spin ferromagnetism, we find in
our case a strong tendency to the formation of Valence Bond (VB) phases. 
In $D=1$, the presence of a statically ordered crystal of VBs is firmly
established. 
Given the presence of a sizeable spin-gap, weakly coupled chains should 
also have a crystal VB ground state. 
In $D=2$ on the square lattice, a spin gap is suggested by our data,
but no sign of crystalline dimer order is found from the relevant
correlation functions. 
A homogeneous liquid of resonating valence bonds appears as the natural 
candidate ground state for this model. 

We thank G. B. Bachelet for his support and useful discussions, and 
M. Calandra for his help. 
Work at SISSA and ICTP was co-sponsored by INFM project HTSC, and
by EU contract FULPROP ERBFMRXCT970155.


%
\begin{center} {\bf FIGURE CAPTIONS} \end{center}

\begin{figure} 
\caption{Finite size gaps for the lowest excited states of 
$H_{\rm ST}$ in $D=1$ (solid symbols) compared to the Heisenberg chain case 
(open squares). 
The excited state for the Heisenberg case extrapolates to 
$\Delta E/J=\pi^2/(2L)+\cdots$, shown by the dashed line. 
The solid straight line, extrapolating to $\approx 0.5969$ for $L\to \infty$
is obtained by a quadratic fit. 
Inset: Log-log plot of $\sum_r (-1)^r \langle S^z_0 S^z_r \rangle$ 
(open squares) and of 
$\sum_r (-1)^r \langle (S^z_0 S^z_{1}) (S^z_r S^z_{r+1}) \rangle$ 
(solid squares). 
The dashed line has slope $1$, for comparison. 
Top: Schematic illustration of the two crystal VB ground states of a linear
chain. }
\label{results_1d:fig}
\end{figure} 
%
\begin{figure} 
\caption{Finite size gap between the ground state and the first excited
triplet state for the ST-model (squares) and for the Heisenberg
model (circles) on the square lattice.\protect\cite{Calandra}  
The data for the ST-model are compatible with the existence of
a spin gap in the thermodynamic limit. 
} 
\label{gap2d:fig}
\end{figure}
%
\begin{figure} 
\caption{ Finite size structure factors obtained by GFMC with the forward 
walking technique. The important components of the Fourier transform of
the dimer-dimer correlation functions look to grow more slowly
than the volume. Therefore we can exclude the presence of any kind of
valence bond solid. For the N\'eel order parameter, 
we find no clear sign of long range order.
The size scaling of the N\'eel order parameter for the Heisenberg model 
is shown for comparison.\protect\cite{Calandra} }
\label{sq2d:fig}
\end{figure}

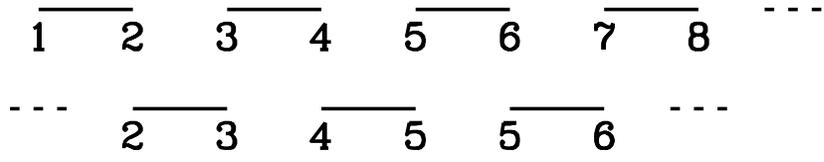
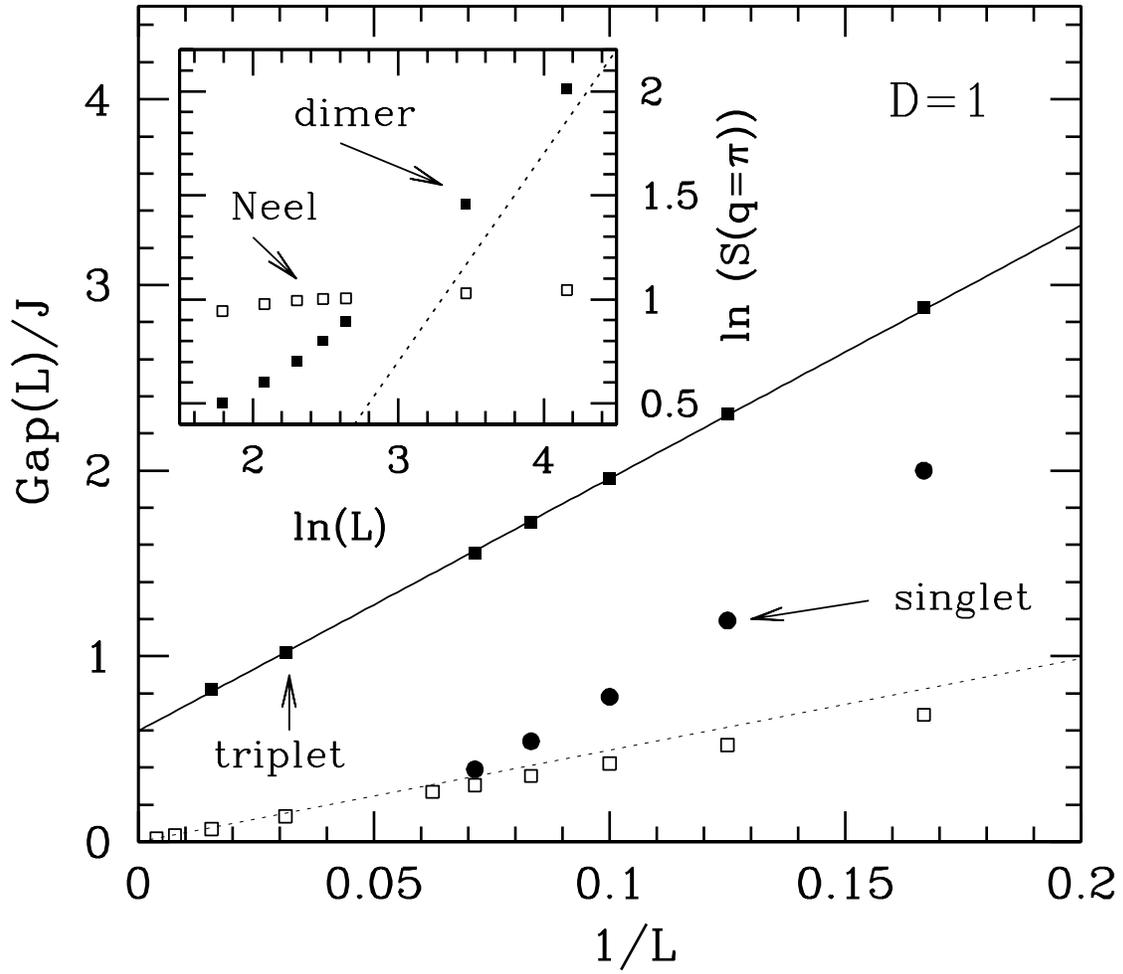
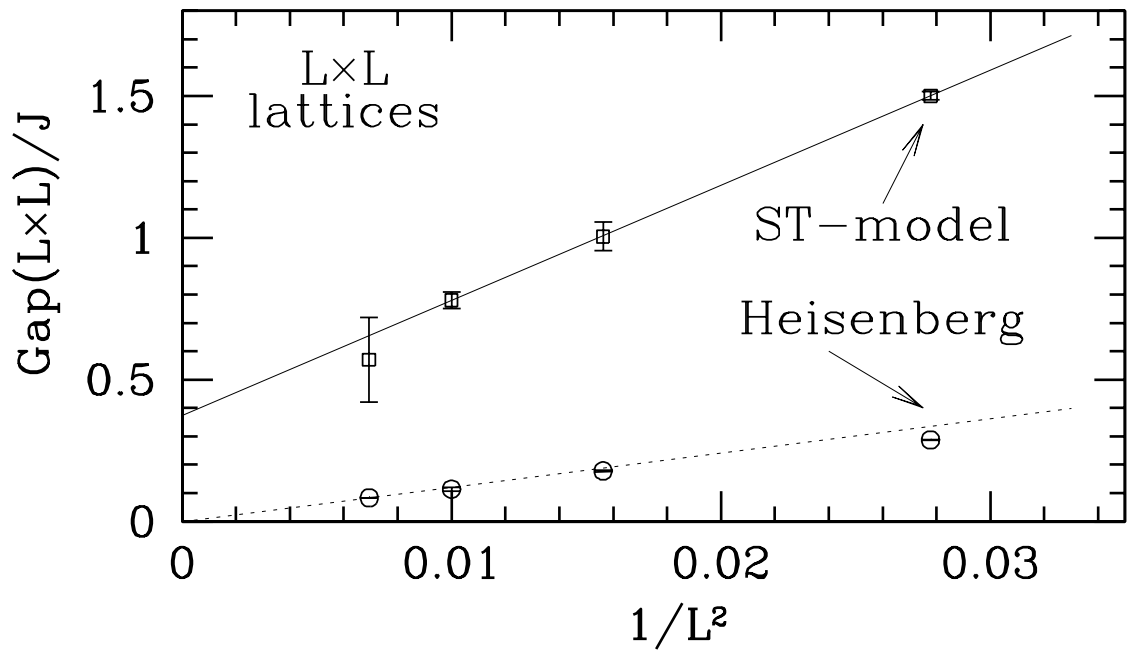
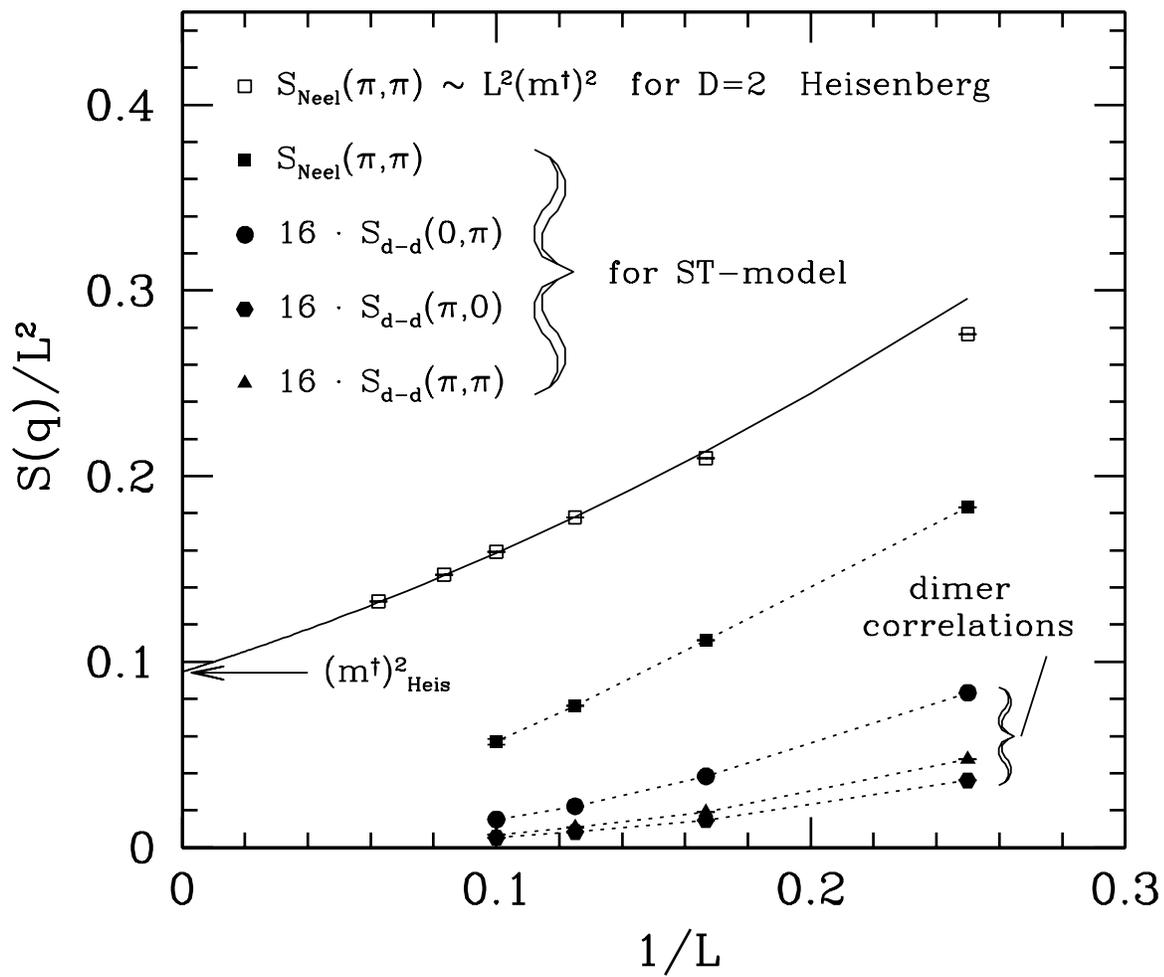
\end{document}